\documentclass{aa}
\usepackage{graphicx}

\usepackage{txfonts}

\begin{document}

\title{The X-ray binary 2S0114$+$650$=$LSI$+$65 010}

\subtitle{A slow pulsar or tidally-induced pulsations?}

\author{G. Koenigsberger\inst{1}, L. Georgiev\inst{2},  E. Moreno\inst{2}, M. G. Richer\inst{3}, O. Toledano\inst{1}, G. Canalizo\inst{4}, 
 \and 
 A. Arrieta\inst{5}
      } 

\offprints{G. Koenigsberger}

\institute{Centro de Ciencias F\'{\i}sicas UNAM, Apdo. Postal 48-3, Cuernavaca, Mor. 62251\\
           \email{gloria@ce.fis.unam.mx; oswaldo@ce.fis.unam.mx}
        \and
          Instituto de Astronom\'{\i}a, UNAM, Apdo. Postal 70-264, Mexico D.F. 04510 \\
           \email{georgiev@astroscu.unam.mx; edmundo@astroscu.unam.mx}
        \and
          OAN, Instituto de Astronom\'{\i}a, UNAM, Apdo. Postal 877, Ensenada, BC, 22800\\
           \email{richer@bufadora.astrosen.unam.mx}
        \and
          University of California, Riverside\\
           \email{gabriela.canalizo@ucr.edu}
        \and
          Universidad Iberoamericana, Mexico D.F.\\
           \email{anabel.arrieta@uia.mx}
         }

\date{Received March 29, 2006; accepted}

\abstract
{The X-ray source 2S0114+650$=$LSI$+$65 010 is a  binary system containing a B-type primary and
a low mass companion believed to be a neutron star.  The system has  three reported
periodicities:  the orbital period, P$_{orb}\sim$11.6 days,   X-ray flaring
with P$_{flare}\sim$2.7 hours  and a ``superorbital"  X-ray periodicity P$_{super}=$30.7 days.}
{The objective of this paper is to show that the puzzling periodicities in the system may be
explained in the context of  scenarios in which tidal interactions drive oscillations in the
B-supergiant star. }
{We   calculate the solution of the equations of motion for one layer of small surface elements 
distributed along the equator of the star, as they respond to the forces due to gas pressure, centrifugal, 
coriolis, viscous forces, and the gravitational forces of both stars, which provides variability timescales 
that can be compared with those observed for 2S0114$+$650.  In addition, we use  observational 
data obtained at the  Observatorio Astron\'omico Nacional en San Pedro M\'artir (OAN/SPM) between 1993-2004 
to determine which  periodicities may be present in the optical region. } 
{The models for circular orbits predict ``superorbital" periods while the eccentric orbit 
models predict strong variations on orbital timescales, associated with periastron
passage.  Both also predict oscillations on timescales of $\sim$2 hrs. We suggest that the 
tidal oscillations lead to a structured stellar wind which, when fed to the neutron star, 
produces the X-ray modulations. The connection between the stellar oscillations and the modulation
of the mass ejection may lie in the shear energy dissipation generated by the tangential motions that
are produced by the tidal effects, particularly in the tidal bulge region. From an observational
standpoint, we find indications for variability in the He I 5875 \AA\ line on $\sim$2 hrs
timescale and, possibly, the ``superorbital" timescale.  However, the line profile variability
exceeds that which is predicted by the tidal interaction model and can be understood in terms of
variable emission that is superposed on the  photospheric absorption.  This emission appears to
be  associated with the B-supergiant's stellar wind rather than the vicinity of the companion. 
}
{The model calculations lead us to conclude that the B-supergiant may be the origin of the 
periodicities observed in the X-ray data,  through a combination of a localized structured  
wind that is fed to the collapsed object  and, possibly, by production of X-ray emission on 
its own surface. This  scenario  weakens the case for 2S0114+650 containing a magnetar descendent.}


\authorrunning{Koenigsberger et al.}
\titlerunning{LSI$+$65 010$=$2S0114$+$650}
\maketitle

\section{Introduction}

The X-ray source 2S0114+650 (Dower \& Kelley 1977) is a  binary system containing a luminous 
B-type star and a low mass companion believed to be a neutron star.  Unlike many such systems, 
however, the short-period X-ray pulsations associated with a rapidly rotating neutron star are
apparently absent.  Koenigsberger et al. (1983) reported $\sim$15 minute pulsations which,
however, were not  found in later observations except those of Ginga (Yamauchi et al. 1990).
On the other hand, X-ray flaring activity was shown to be periodic
(Finley et al. 1992;  Hall et al. 2000) with P$_{flare}\sim$2.78 hrs, a period that
in more recent INTEGRAL data has decreased to $\sim$2.67 hrs (Bonning \& Falanga 2005). 
Finley et al. (1992) suggested that this period could actually be due to the pulsar's rotation 
which, however, would make it an extremely slow pulsar.  Li \& van den Heuvel (1999) 
showed that the  neutron star in the 2S0114+650 system could  have reduced its rotation rate 
within the lifetime of the B-supergiant companion only if it had a very large 
($>10^{14}$ Gauss) magnetic field at the time of its birth. Thus, the new-born 
pulsar would have been a {\it magnetar}. 
  
An alternative scenario for explaining the  2.7 hour periodicity that was also suggested
by Finley et al. involves  pulsations of the optical component by which a structured
stellar wind with alternating high and low density regions is produced. The   density 
variations of material being accreted by the collapsed companion would lead to the
observed periodic flaring activity.  However,  it is not evident: a) whether a B-supergiant
may oscillate with this short period;  nor b) how the stellar oscillations  can produce the 
variable mass-loss;  nor c) how the ejected mass can be  distributed in the precisely structured 
stellar wind that is needed for the periodic modulation of the accretion rate onto the neutron star.  
If a plausible scenario involving pulsations of the B-star could be constructed, the {\it magnetar} 
hypothesis would be weakened. 

A second  X-ray variability timescale in 2S0114+650 is the orbital period.  Corbet et al.
(1999) suggested that this variability may be due to eclipses of the X-ray source,  but
Mukherjee \& Paul (2006) find that the change in the ASCA X-ray luminosity can
be ascribed to a local change in the density of the stellar wind from which the
neutron star accretes.  The latter would occur  either if the orbit is 
eccentric  or if the mass-loss characteristics of the B-star are orbital-phase
dependent.  Interestingly, the description given by  den Hartog et al. (2006) of the
broad maximum in the RXTE-ASM and INTEGRAL data as consisting mostly of flares supports
the idea of a non-steady mass-loss rate from the B-star, with larger amplitude variability
at phases when the average X-ray flux is maximum.  This behavior is one that might be expected
in eccentric orbits.  On the other hand, however, a third periodic X-ray variability timescale 
has been recently reported by Farrell et al. (2006): P$_{super}\sim$30.7 days, which they
suggest may be caused by a warped accretion disk around the collapsed object.  However,
the absence of He II 4686 \AA\ emission (Reig et al. 1996; Koenigsberger et al. 2003) 
argues against the presence of a significant accretion disk. 


Non-synchronously rotating stars in binary systems are well-known for exhibiting 
numerous periodicities due to the oscillations that are driven by the tidal forces (Press \& Teukolsky 
1977; Zahn 1977; Savonije et al. 1995; Kumar et al. 1995). In particular, Moreno et al. (2005)
found ``superorbital" periodicities in models of binary systems with circular orbits, in
addition to shorter period variability. Thus, the question naturally arises 
of whether   we can produce a consistent picture for explaining  the puzzling periodicities 
of 2S0114+650 through scenarios involving tidal interactions.

In Section 2 we present the observational data;  in Section 3 we summarize the tidal 
interaction model that we use and  describe its predictions for an eccentric 
and a circular orbit;  in Section 4  we describe  the behavior of the He I 5875 \AA\
line; and  Section 5 contains  our conclusions.



\begin{table*}
\caption{Radial velocity and residual emission He I 5875 \AA}
\label{table:1}
\centering
\begin{tabular}{lllllllrlr}
\hline\hline
Epoch& Instrum.& $<MJD>$ & N & S/N &
$\phi_{orb}$& $\phi_{sup}$ & s.d.(ISM)& RV ($\pm$s.d.)& $-$EW  \\
     &         &       & &   &     &        & km/s     & km/s   &    \AA        \\
\hline
1993Oct& REOSC   & 49269.54& 5  & 26(8)  &  0.804&0.78  & 1.1 & $-$41.5 (1.8) & 0.35\\
         & --- & 49270.33& 5  & 29(7)  &  0.881&0.80   & 1.5  &$-$37.7 (2.3)& 0.39  \\
         & --- & 49273.39& 14 & 20(7)  &  0.145&0.90   & 2.6 & $-$41.7 (3.9)& 0.45 \\
1995Nov& UH coud\'e  & 50049.38& 13 &36(5)   &  0.052&0.11   & 0.7 &$-$65.8 (4.6) & 0.61 \\
       & --- & 50050.34&  9 &33(9)   &  0.135&0.14   &  0.4 & $-$63.6 (3.5)&0.65   \\
2001Jan& REOSC   & 51931.18& 11 &84(17)  &  0.304&0.22   & 2.1& $-$53.9 (3.8) & 0.00 \\
2001Oct& REOSC   & 52195.36& 12 &68(11)  &  0.082&0.80   & 3.4 & $-$36.9 (4.2)& 1.10  \\
2003Jan& MES     & 52657.40& 12 & 38(5)  &  0.920&0.81   &1.3 & $-$50.3 (3.3) & 1.09 \\
2004Oct& MES     & 53307.33& 2  & 60     & 0.957&0.92   & ---  & $-$23.3 & 0.60       \\
       & --- & 53308.30& 2  & 50     & 0.041 &0.95  & --- & $-$36.5 &  0.99      \\
       & --- & 53309.27& 1  & 53     & 0.125 &0.98  & ---   & $-$44.0 & 0.97       \\
2004Nov& REOSC   & 53311.34& 15 & 60(11) & 0.304 &0.05  &  2.4 & $-$62.3 (5.2)& 0.70  \\
       & --- & 53314.30& 3  & 55     & 0.559 &0.14  & ---    &$-$87.6 &  0.87     \\
       & --- & 53315.20& 2  & 62     & 0.637 &0.17  & ---    &$-$77.4 &  0.59     \\
       & --- & 53316.30& 2  & 47     & 0.731 &0.21  & ---    &$-$75.0 &  0.81     \\
2004Nov& MES     & 53328.16& 1  & 69     & 0.754&0.59    & ---    & $-$47.5& 0.75         \\
       & --- & 53329.43& 1  & 74     & 0.864 &0.64  & ---    & $-$42.3& 0.66        \\
       & --- & 53330.29& 1  & 53     & 0.938 &0.66  & ---    & $-$38.8& 0.57        \\
       &  ---& 53333.40& 2  & 70     & 0.206 &0.77  & ---    & $-$63.2& 0.62        \\
       & --- & 53334.41& 1  & 60     & 0.293 &0.80  & ---    & $-$70.7& 0.21        \\
       & --- & 53335.40& 1  & 66     & 0.378 &0.83  & ---    & $-$80.2& 0.49        \\
2004Dec& MES     & 53354.23& 16 & 58(9)  & 0.002 &0.44  & 1.5 & $-$30.1 (1.3) & 0.84    \\
       & --- & 53355.20& 17 & 53(6)  & 0.085 &0.47  & 1.0 & $-$43.6 (1.5)& 0.90  \\
       & --- & 53356.20& 15 & 53(13) & 0.172 &0.51  & 1.1 & $-$46.0 (1.5)& 0.77  \\
       & --- & 53357.20& 18 & 54(8)  & 0.258 &0.54  & 1.1 & $-$59.1 (2.1)& 0.95  \\
       & --- & 53358.20&  8 & 59(9)  & 0.344 &0.57  & 0.3 & $-$60.8 (1.6)& 1.15  \\
\hline
\end{tabular}
\end{table*}
\section{Observations and data reduction}

Nine different epochs of observations of LSI+65 010 are analyzed  in this paper. 
The data of four of these  were described in Koenigsberger et al. (2003) three of which 
(1993 Oct 9, 10, 13;  2001 Jan; and 2001 Oct) were obtained with the REOSC echelle spectrograph on the 
2.1 m telescope of the Observatorio Astron\'omico Nacional en San Pedro M\'artir (OAN/SPM) 
and one  was obtained in 1995 with the coud\'e spectrograph on the University of Hawaii 2.2 m 
telescope.  The new sets of data  were obtained with the 2.1m OAN/SPM telescope as follows: 
on UT 2003 Jan 18, 2004 Oct 29-31,  Nov 19-26, and  Dec 15-19 with the MES-SPM 
spectrograph,  and on 2004 Nov. 2 and Nov. 5-7 with the REOSC echelle  spectrograph. 
The total number of spectra that were analyzed is 182.  

The REOSC echelle camera system is described by Diego \& Echevarria (1994). 
With the 300 grooves mm$^{-1}$ echellete, the system gives a resolution  of of 
7.8 \AA\ mm$^{-1}$ (0.15 \AA\ pixel$^{-1}$) at H $\beta$ and 10.7 \AA\ mm$^{-1}$ 
(0.20 \AA\ pixel$^{-1}$) at H$\alpha$. The resolution is R=17,000, which corresponds  
to 17 km s$^{-1}$ per 2 pixels.  The slit was set at a width of 150 $\mu$m, which 
corresponds to 2\arcsec.  The camera of the MES-SPM spectrograph is of a slow focal 
ratio, f/7.5,  producing a stability in the wavelength dispersion 
of $\pm$0.03 \AA\, or a velocity dispersion of $\pm$1.5 km s$^{-1}$  (Richer, 2003).
LSI+65 010 was observed through the He I 5875 \AA\ filter, which covers a 
bandwidth of approximately $\lambda\lambda$5860-5903 \AA\ allowing the use of the
interstellar Na I D  lines at $\lambda\lambda$5889, 5895 \AA\ as  absolute 
velocity standards.  Spectra of LSI+65 010 were interleaved with spectra of the Th-Ar lamp. 

 
Data reduction was performed with IRAF\footnote{IRAF is distributed by the National 
Optical Astronomy Observatories, which are operated by the Asociation of Universities 
for Research in Astronomy, Inc.,under cooperative agreement with the National 
Science Fundation.} and MIDAS\footnote{MIDAS is the acronym for the  Munich Image 
Data Analysis System which is developed and maintained by the European Southern 
Observatory.} using  standard reduction procedures.  Table 1 summarizes the general 
characteristics of the  data sets. Column 1 identifies the epoch of observation, column 2 
the instrument, column 3 lists the average Julian date, column 4 the number of spectra, and 
column 5 lists the average signal-to-noise ratio and standard deviation about the mean 
(in parenthesis). This is the average  S/N ratio of the individual spectra. The standard
deviation is given only for nights when at least five individual exposures were made. 

The MES-SPM instrumental response has a very complicated shape, due to the interference
filter used, with a bump very near the red edge of the He I line.  If not taken into 
account properly, this bump could be interpreted as
a red emission component.  In order to avoid this problem, the spectrum of the radial
velocity standard star HD22484 was used to rectify the
continuum.  This was done by cutting out the photospheric lines, which are all very
narrow in this star,  and the resulting ``cleaned"  continuum distribution was used
to normalize the spectra of LSI+65 010.

The stability of the wavelength calibration  was tested by measuring on each individual
spectrum  at least one of the interstellar lines in the vicinity of He I 5875.80 \AA. 
The Na I $\lambda$5889, 5895 \AA\ ISM lines in LSI+65 010 have  several velocity components 
(Koenigsberger et al. 2003), two of which are narrow and clearly resolved in the UH-coud\'e
and MES-SPM data. We used the highest velocity component of $\lambda$5889.95 \AA\, located 
at $\sim$5888.90 \AA\, to establish the precision of the velocity measurements for these
spectra.  This ISM component is not resolved in the echelle data.  Hence, we used the zero-velocity
component of Na I 5889.95 \AA\ and the  diffuse interstellar band (DIB; Herbig 1995) located at 
$\lambda$5849.887 \AA\ to determine the precision of velocity measurements on the REOSC spectra.
Column 8 of Table 1 lists the standard deviation about the average value of the chosen ISM 
features, for all cases in which 5 or more spectra were obtained.  In general, the precision of the
 velocity measurements in the UH-coud\'e and MES-SPM spectra is better than 1.5 km s$^{-1}$, while
the REOSC echelle data allows RV measurements with a precision better than $\sim$4 km s$^{-1}$. 

In order to account for possible zero-point shifts in the wavelength scale, and at
the same time, correct for the motion of the Earth and Sun, all radial
velocities of He I $\lambda$5875 \AA\ were measured with respect to the centroid of the 
nearby interstellar absorption lines.
In practice, this was accomplished by: a) fixing v=0 km s$^{-1}$ on the  velocity scale at either
$\lambda$5849.89 \AA\, $\lambda$5888.90 \AA\  or $\lambda$5889.95 \AA; b)  measuring the velocity
of the ISM component and the He I 5875 photospheric line; c) subtracting the measured velocity
of the ISM line from the He I line velocity.   This procedure was followed for each
individual spectrum.  The ISM lines were measured by fitting one or several Gaussian functions
while the centroid of the He I 5875 \AA\ line  was measured with IRAF by
integrating the pixel values between the two edges of the absorption line profile
that intersect the level  0.9 of the normalized continuum.  This method for He I 5875 \AA\  was
chosen primarily because it reduces the effects produced by the variability of the 
extended wings of the line.
Column 9 of Table 1 lists the average value of the  RV, obtained by averaging the  
values of the individual RV measurements of He I corrected for the shift in the ISM line, 
for each night ($\pm$s.d. in parentheses).  This velocity is with respect to the  laboratory 
wavelength, adopted as $\lambda_0=$5875.80, which is the average wavelength of the two
tabulated He I lines at $\lambda\lambda$5875.62 and 5875.97 \AA.  

\section{The tidal interaction model}

Moreno \& Koenigsberger (1999) and Moreno et al. (2005)  presented a simple model that
describes the behavior of a star's surface layer in the presence of the
perturbing effect of a binary companion.   The model calculates the solution of the 
equations of motion for one layer of small surface elements distributed along the equator 
of the star as they respond  to gas pressure,  centrifugal, coriolis  and viscous forces, 
as well as the gravitational forces of both stars.  The equations of motion are solved 
in the non-inertial reference frame that is centered on M$_1$, the star being modeled, and 
that rotates with the same angular velocity as the orbital motion of M$_2$, the companion star. 
The main body of M$_1$, interior to the thin surface layer, is assumed to behave as a rigid body, 
with the tidal deformation appearing only in the surface layer.   

For the numerical calculation, the required input parameters are M$_1$ and M$_2$, M$_1$'s radius 
and rotation velocity, R$_1$ and v$_{rot}$, respectively, 
the orbital period P$_{orb}$, the eccentricity $e$, the inclination of the orbital plane {\it i},
the argument of periastron $\omega_{per}$, the viscosity $\nu$, and the relative thickness 
$\delta$R/R$_1$ of M$_1$'s  external oscillating layer.  The output  consists of temporal 
series describing, among other characteristics, the maximum perturbation in  stellar radius, 
R$_{max}$,  and radial velocity, V$_{max}$, the azimuthal component of the velocity perturbation 
for all azimuth angles at given points in time, $\Delta$V$_{\Phi '}$, and the photospheric absorption 
line profiles as detected by an observer in an external inertial reference system.  

The  orbital eccentricity and the stellar rotational velocity are  two critical
parameters for understanding the variability patterns in binary systems.  If e$>$0, 
the perturbation produced during periastron passage is a dominant effect and hence, the strongest 
variability occurs on orbital timescales.  On the other hand, if e$=$0, ``superorbital" or
``suborbital" periodicities appear.  Whether the periodicity is ``superorbital"
or not depends on the ratio of the rotational angular velocity to the orbital angular 
velocity, $\beta_0=\omega_0/\Omega_{per}$. Here $\omega_0=$2$\pi$/P$_{rot}=$v$_{rot}$/R$_1$  
is the  rotation angular velocity of the  rigidly rotating inner stellar region, and R$_1$ the 
unperturbed surface stellar radius. $\Omega_{per}$  is the orbital angular velocity, and its value
at periastron is adopted for the general case of an eccentric orbit.  
Thus, $\beta_0$ is a measure of the non-synchronicity between the stellar rotation 
and the orbital angular velocities at periastron, and it can be written as:
\begin{equation}
\beta_0=0.02 \frac{P_{orb} v_{rot}}{R_1}\frac{(1-e)^{3/2}}{(1+e)^{1/2}} 
\end{equation}

\noindent where $v_{rot}$ is in units of km s$^{-1}$, R$_1$ in units of R$_\odot$, P$_{orb}$ in days, 
and {\rm e} is the eccentricity of the orbit.


\subsection{Application to LSI$+$65 010}

The spectral line profiles of LSI$+$65 010 have peculiar shapes and are variable (Koenigsberger et 
al. 2003) which results in a relatively large uncertainty for  
{\rm v}~{\it sin i}=96$\pm$20 km s$^{-1}$ (Reig et al. 1996). We illustrate in Figure 1 the 
theoretical rotationally broadened 
line profiles for {\rm v}~{\it sin i }=96 km s$^{-1}$ and 120 km s$^{-1}$. The cores of
both of the  observed profiles are indeed well matched with the {\rm v}~{\it sin i }=96 km s$^{-1}$
theoretical prediction, but their blue wings are significantly more extended.  For comparison,
the observed line profile obtained by Canalizo et al. (1995) of the B-supergiant optical counterpart 
of Cyg X-1, HDE 226868, is also included in this figure.   HDE 226868's absorption is better matched by 
the {\rm v}~{\it sin i }=120 km s$^{-1}$ theoretical profile,  although the extent of its blue wing 
is also underpredicted.  The blue wing of strong photospheric absorptions in early-type stars  
is generally  more extended due to the expansion of the external atmospheric layers as they accelerate and  
become the base of the stellar winds.  Variability in the blue wing as displayed in Figure 1 for LSI$+$65 010 
indicates that the atmospheric properties at the base of the wind undergo significant time-dependent changes. 
 These changes occur on timescales of $\sim$days (Koenigsberger et al. 2003).

\begin{figure}
\centering
\includegraphics[width=9cm]{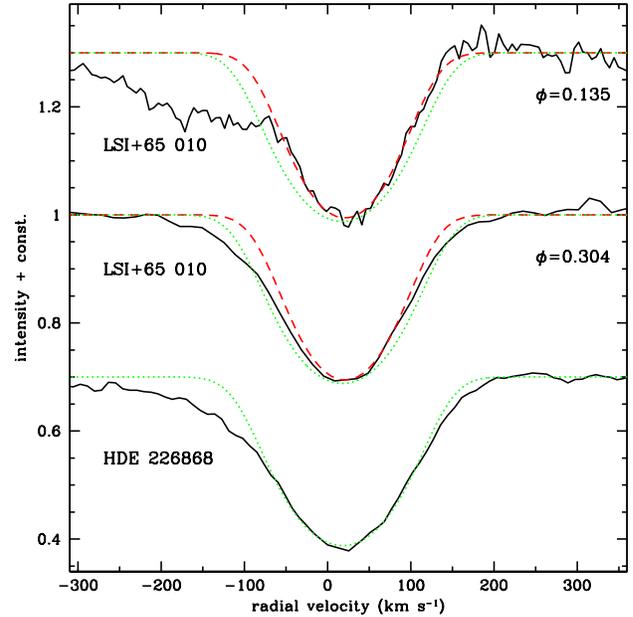}
\caption{Observed line profiles of LSI$+$65 010 at two different orbital phases
compared with theoretical lines profiles rotationally broadened to v sin(i)=96 km s$^{-1}$
(dashes) and 120 km s$^{-1}$ (dots), and with HDE 226868, the  B-supergiant counterpart
of Cyg X-1. Profiles are shifted by $\pm$0.03 on the vertical scale for clarity in the figure.
}
\end{figure}

It is important to note that LSI+65 010's  profile at $\phi=$0.304 shown in Figure 1   
substantially resembles  HDE 226868's line profile, as well as the theoretical profiles.  
Thus, the $\phi=$0.304 spectrum  corresponds to a time when the B-supergiant is minimally
perturbed and is probably the best profile for estimating {\rm v}~{\it sin i }. 
Adopting the value found by Reig et al. (1996) and the reported estimates of {\it i }$\sim$61$-$90$^\circ$,
v$_{rot}\sim$96$-$109 km s$^{-1}$.  The other stellar parameters derived  
for LSI+65 010 are M$_1=$16$\pm$5 M$_\odot$, R$_1=$37$\pm$15 R$_\odot$, and  
M$_2=$1.7 M$_\odot$ for the neutron star (Reig et al. 1996).  

Keeping M$_1=$16 M$_\odot$, M$_2=$1.7 M$_\odot$ and e=0 fixed, we performed a series of model 
calculations searching for variability on the  P$_{super}\sim$30.7 days,  finding that  
P$_{super}$ increases linearly with $\beta_0$ for $\beta_0$ in the range 0.57$-$0.63.  
P$_{super}\sim$30.7 days was achieved with $\beta_0=$0.615.   Fixing this value in Eq. (1)
constrains  R$_1$ to the range 36.2$-$41.1 R$_\odot$, for values of v$_{rot}$ between 96$-$109 
km s$^{-1}$.  Figure 2 illustrates the characteristics of the ``superorbital cycle" for two parameter 
sets (R$_1$,v$_{rot}$)=(40 R$_\odot$, 106 km s$^{-1}$) and (36.2 R$_\odot$, 96 km s$^{-1}$).   
A slow increase in the size of the tidal bulge over most of the cycle is followed by
a more abrupt growth to an absolute maximum lasting $\sim$5 days.  Oscillations on timescales
of $\sim$hours also appear and, for the cases illustrated in Figure 2, they lie in the range 2-3 hours.

\begin{figure}
\centering
\includegraphics[width=9cm]{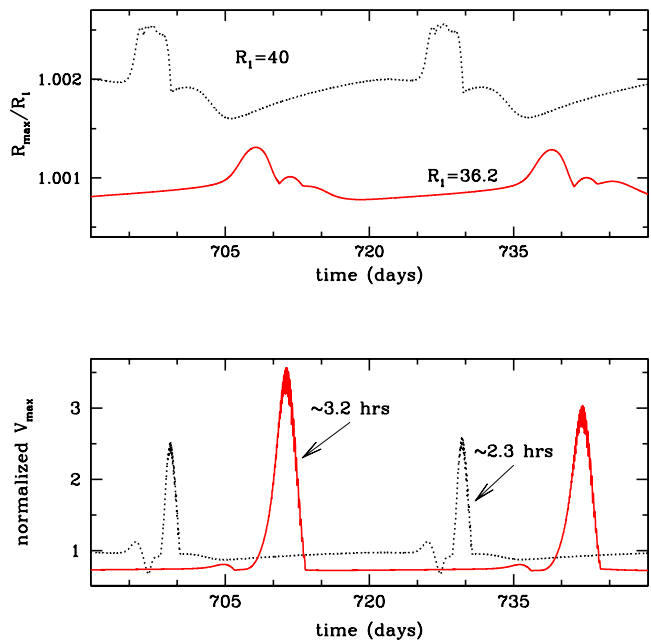}
\caption{Radial extent of the tidal bulge, R$_{max}$ (top), and maximum
surface velocity, V$_{max}$, plotted as a function of time for  circular
orbit models with $\beta_0=$0.615. Results for two different stellar radii,
R$_1=$40 R$_\odot$ (dotted line; model C-1) and R$_1=$36.2 R$_\odot$ (model C-2) 
are shown, illustrating the $\sim$31-day ``superorbital" period  that appears. 
Superposed oscillations on $\sim$hours timescales are 
indicated by the arrows, though they are barely visible on this figure.
}
\end{figure}

In contrast to the circular orbit case, there are no ``superorbital" periods on $\leq$100 day-timescales
in eccentric orbit case, so the value of $\beta_0$ is  unconstrained except for the limits on 
v$_{rot}$ and R$_1$ given by Reig et al., which lead to the range $\beta_0\sim$0.40-0.90.
An example of the tidal interactions predicted by the model for $\beta_0=$0.66 is shown in Figure 3.
The data for one orbital cycle are plotted and  the abscissa is given in units of
orbital phase with periastron ($\phi=$0.0) and apastron  indicated by the vertical tick marks.   
It is interesting to note that although the  tidal bulge grows gradually as periastron is approached, 
it reaches  maximum extent only {\em after} periastron passage due to the retarding
action of the viscosity.

The short-timescale oscillations  are most clearly seen in the bottom plot of Figure 3. 
Their frequency gradually decreases with time and corresponds to periodicities in
the range $\sim$1.6-3.4 hours.   The predicted amplitude is largest in the phase
interval $\sim$0.15-0.40.  

The idea of tidal oscillations in the tangential direction is somewhat less familiar
than that of the radial oscillations.  However, the ``swashing" back and forth of the surface
elements as they approach and pass the sub-stellar point is responsible for some of
the strongest variability that is produced in photospheric line profiles. The  tangential 
velocity perturbation, $\Delta$ V$_{\Phi '}$ for the e$=$0.16 case is illustrated in Figure 4
for different orbital phases, showing that the largest amplitudes occur near the primary tidal
bulge, $\Phi '\sim$360$^\circ$.  Note also that the perturbations are  strongest {\em after}
peristron passage, around orbital phase 0.20.

In Table 2 we list the observationally-derived parameters of LSI$+$65 010 as well as the 
parameters of the model calculations for the eccentric orbit and the two calculations for the
circular orbit.


\subsection{Viscosity and shear energy dissipation}
The R$_1=$40 R$_\odot$ models are highly unstable, which is easy to understand considering that
this radius is very close to the Roche radius,
requiring relatively
large values of the viscosity to allow the computation to be performed. The oscillation
amplitudes scale inversely with the viscosity, so larger amplitudes would  be attained if
we could use smaller $\nu$ values. However, as described by Moreno et al. (2005), when the 
oscillation amplitudes become too large, the individual surface elements for which the 
equations of motion are being solved become detached or begin to overlap, at which time the 
numerical computation is halted. 
To prevent this from happening, the value of $\nu$ may be increased.  Our practice is to
choose the smallest value of $\nu$ that allows the code to run over at least $\sim$5 orbital
cycles in order to compute the line profiles after the initial transitory phase of the
calculation.  In the real star, the
phenomena that we describe as ``overlapping or detached surface elements" have no
significance except that they suggest the possibility of large-scale relative motions that
may be interpreted in terms of the more familiar term {\it macro-turbulence}.  With this
interpretation in mind, the value we use for $\nu$ in the calculations may be  thought
of as describing the conditions of the stellar surface at the limit where large-amplitude 
turbulent motions develop on small length scales.  

The viscosity is related to the $\alpha$-parameter (Shakura \& Sunyaev 1973) used in studies 
of accretion disks in Keplerian orbits through $\nu\sim(\alpha/\omega$)({\it k}T/$\mu$m$_H$), 
where T is the temperature, $\mu$ is the mean molecular weight, and $\omega$ is a characteristic 
angular velocity.  Associating $\omega$ with the relative  motion of the external layer 
with respect to the rigidly-rotating inner region of our model, and assuming T=20,000 K,
the values of $\nu$ we use in this paper all correspond to  $\alpha<$0.01, well within the 
range expected  for the outer layers of a supergiant star.

\begin{figure}
\centering
\includegraphics[width=9cm]{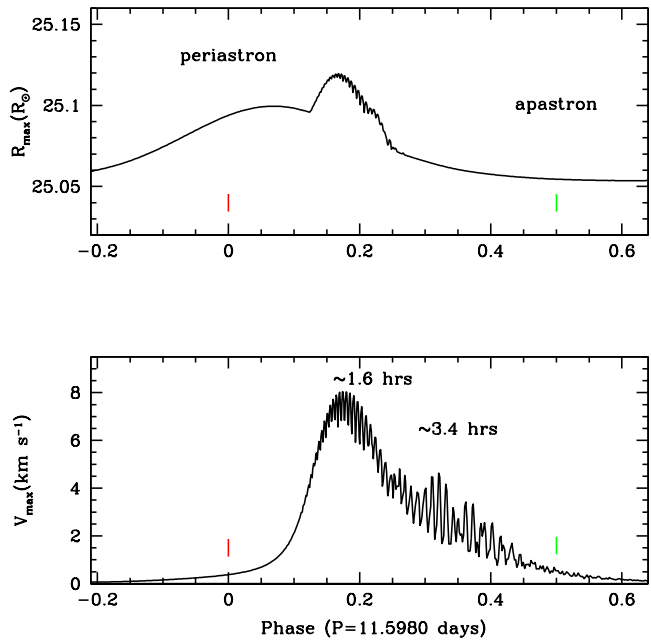}
\caption{Radial extent of the tidal bulge, R$_{max}$, and maximum
surface velocity, V$_{max}$, over one orbital cycle for the e=0.16 eccentric
orbit.  The abscissa is orbital phase.  Periastron ($\phi=$0.0)
and apastron  are indicated. The short-timescale oscillations, best
seen in the bottom panel, have periods $\sim$hours that gradually increase after periastron.
}
\end{figure}




A direct consequence of  the behavior of $\Delta V_{\Phi '}$ and of including the viscosity
in the calculation is that the surface layer sliding over the inner, rigidly-rotating body 
leads to shear energy dissipation, $\dot{E}$.  Note that  $\dot{E}$ is a potential source of
additional energy that may be fed into the surface layers of the star and it has been suggested 
that $\dot{E}$ can contribute towards driving stellar winds or mass-ejections,
or may play a role in the generation of magnetic fields (Koenigsberger et al. 2002; Moreno et al. 2005; 
Toledano et al. 2006).  All these processes are generally associated with X-ray emission.  Within 
this context, Figure 4  implies a concentration of  surface activity on the 
hemisphere of M$_1$ that faces its companion. Hence, in addition to the effects produced by
the accretion onto the neutron star and the ensuing X-ray emission, a strong degree of variability due
to the activity on the B-star surface facing its companion is expected, particularly at periastron
passage (if the orbit is eccentric). 

\begin{figure}
\centering
\includegraphics[width=9cm]{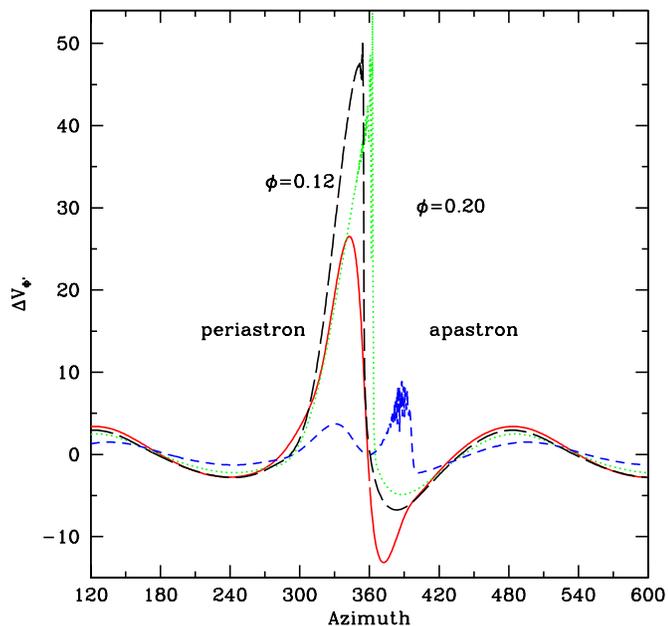}
\caption{Azimuthal velocity perturbation of the surface layer as a
function of angle for the eccentric orbit case for four different orbital phases,
as indicated.  Note that the largest azimuthal velocities occur near $\phi=$0.12,
rather than at periastron passage. Large values of $\Delta$ V$_{\Phi '}$  lead  
to large energy dissipation rates.
}
\end{figure}

\begin{figure}
\centering
\includegraphics[width=9cm]{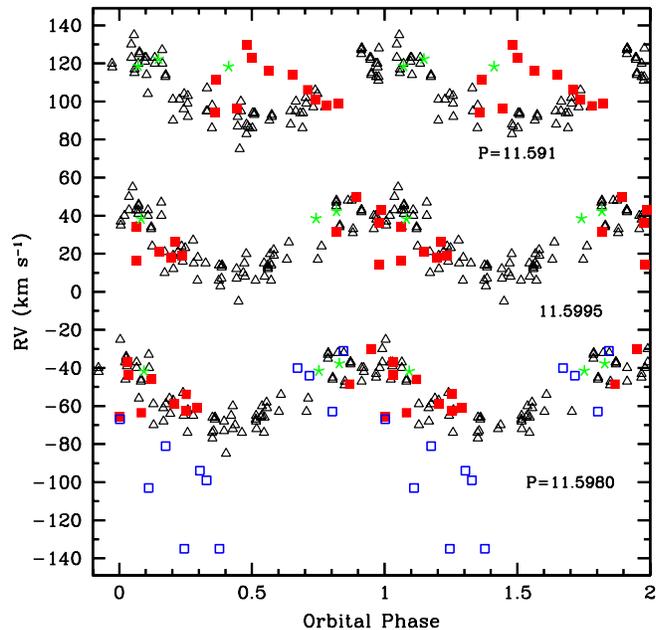}
\caption{Radial velocity curves for three different periods: Crampton et al.'s (1985) 11.591 (top, shifted
by +120 km s$^{-1}$), den Hartog et al.'s 11.5995 (middle, shifted by +60 km s$^{-1}$) and
11.5980 days (bottom). Data plotted are: Crampton et al. (1985)
(open triangles),  Reig et al. (open squares), our 1993 data (stars) and our 1995-2004 averages
(filled squares) of at least 5 spectra per night.
}
\end{figure}

\begin{figure}
\centering
\includegraphics[width=9cm]{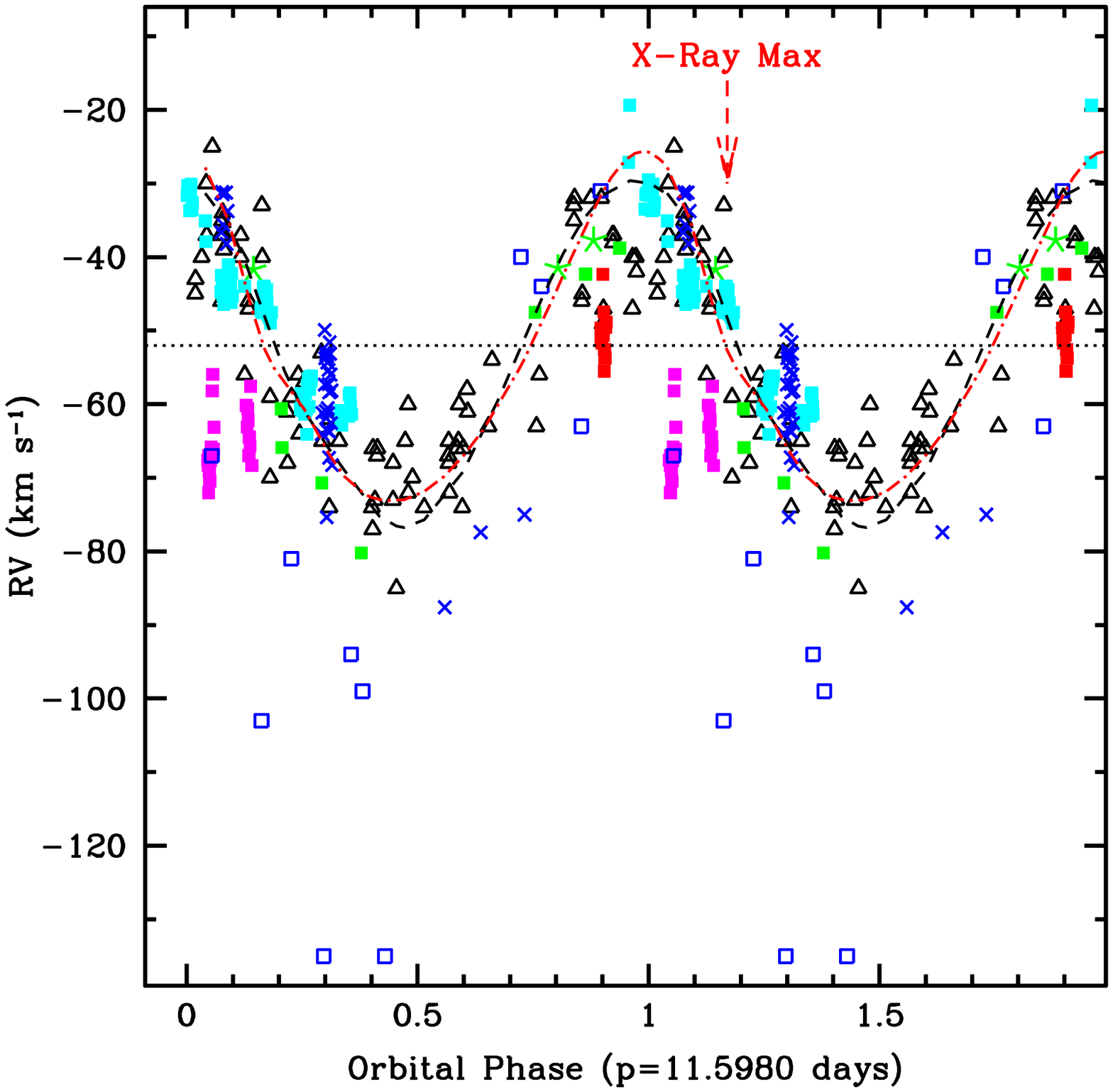}
\caption{Radial velocity of  Crampton et al. (open triangles), Reig et al. (open squares), and our 
measurements on individual spectra: REOSC echelle (crosses), UH coud\'e and MES-SPM (filled squares). 
The curves represent the  RV that would be measured from of a M$_1=$16 M$_\odot$ primary star tidally
perturbed  by a M$_2=$1.7 M$_\odot$ companion in a $e=$0.16 (dot-dashes) or a circular (dashes)
orbit.  Phase 0.0 is maximum receding velocity for the circular orbit and periastron passage
for the eccentric one. Time of X-ray maximum according to  dem Hartog et al. ephemeris but
with P=11.5980 days is indicated.
}
\end{figure}

The two approximations  used in our model calculations that must be noted are the artificial
division of the star into an outer oscillating layer and an inner rigidly rotating body,  and
the fact that the calculation of the oscillations is performed only for the equatorial
region of the star. The former limits the ability of the code to reproduce many of the oscillation
frequencies that depend on the structure of the entire star, while the latter yields upper limits
to the oscillation amplitudes since the strongest perturbations occur in the orbital plane. 
Finally, we note that the line-profile calculation does not currently include effects due
to possible variations in the effective temperature over the stellar surface.

\section{LSI+65 010: Results from the observational data}

\subsection{RV curve}

The B-type optical counterpart of 2S0114+650, LSI+65 010=V662 Cas, was
first identified by Margon \& Bradt (1977) based on its variable H$\alpha$ emission and
classified by Reig et al. (1996) and Liu \& Hang (1999) as  B1 Ia.  Using measurements
of photospheric absorption-line radial velocity (RV) variations from data obtained
between 1978 and 1984, Crampton et al. (1985) concluded that the orbit of the B-supergiant
was either circular, with P$_{orb}^{circ}=$11.591$\pm$0.003 days or slightly eccentric
(e$=$0.16),  with P$_{orb}^{ecc}=$11.588$\pm$0.003 days.   Recently, den Hartog et al. (2006)
found P$_{orb}=$11.5995$\pm$0.0053 days from an analysis of  RXTE-ASM and INTEGRAL data. 

Figure 5 presents a plot (top) of the Crampton et al. (1985)  data and our average data from 
Table 1 all folded in phase with the  circular-orbit period and initial epoch.
Our data in this figure are limited to points for  which  5 or more individual spectra 
were obtained during the night.
A significant phase shift between the modulation of the Crampton et al. (1985) data and the more
recent data is evident. This phase shift  is reduced  if the data are folded 
with the den Hartog et al. P$=$ 11.5995 days (middle plot in Figure 5).  
An even better agreement between the two data sets is achieved with
P$=$ 11.5980 days, which is well within the uncertainties given by  den Hartog et al. 
The bottom plot of Figure 5 also includes the data published by Reig et al. (1996),
obtained from RV measurements of He I 6678 \AA.  They follow in part the general
trend of the other two data sets, but display  significantly
larger negative RV at phases $\sim$0.1-0.4 which correspond to orbital phases in
which the secondary star is ``in front" of the B-supergiant.  This means that the
photospheric absorption lines must have had a significantly more
extended blue wing during these orbital phases, suggesting a difference between the 
hemisphere of the B-star that faces its companion and the opposite hemisphere.  
Considering the discussion of the previous section, we speculate that the difference 
between the two hemispheres may be associated with the tidal bulge and its oscillations.

In Figure 6 we now plot all the individual data measurements folded with P=11.5980 days
and T$_0=$43669.9 MJD.  This value of T$_0$ is derived by taking  the initial epoch given
by Crampton et al. (1985) for their circular orbit solution and subtracting 40 cycles of 11.5980
days each.  T$_0$ corresponds to the time of fastest {\it receding} radial velocity, which
means that the hemisphere of the B-supergiant that faces its companion starts coming into
view at $\phi=$0 and is occulted in our line-of-sight at $\phi>$0.5.
The initial epoch T$_{ecc}$  of the Crampton et al. (1985) eccentric orbit solution defines
phase 0 as periastron passage.  The different initial epochs are all listed in
Table 2.   We indicate with an arrow in Figure 6 the center of the broad X-ray maximum reported
by den Hartog et al.  According to their  ephemeris, the  X-ray ``high" state is centered
at $\phi\sim$0.12 for the circular orbit solution or $\phi\sim$0.07 if the orbit is
eccentric.  It is difficult to understand why X-ray maximum should be centered at these phases
instead of around $\phi=$0.25 unless the orbit is eccentric, in which case X-ray maximum
would occur very close to the phase for which our tidal model predicts maximum perturbation
of the tidal bulge (see Figure 2).  

The curves  drawn in Figure 6 correspond to the two orbital solutions (e$=$0, e$=$0.16) but,
compared to the large scatter of the data, are nearly indistinguishable. As we will show below, 
the dispersion of the RV data is intrinsic to the line profiles. Thus, the RV's contain 
information additional to the orbital motion making the determination of the orbital solution
very uncertain.  It is interesting to note that the RV curves of other HMXRBs 
such as HD 153919/4U1700-37 (Hammerschlag-Henseberge et al. 2003)
and  HD77581/Vela X-1 (Barziv et al. 2001) also suffer from large intrinsic scatter,
introducing significant uncertainties in the determination of the masses of their respective
collapsed companions.  

The individual RV measurements made on the MES-SPM data are plotted in Figure 7, showing 
that there is a systematic trend for cyclical variability with a characteristic timescale of
$\sim$2.1 hours. The error bars correspond to the standard deviation of ISM line 
measurements, and are significantly smaller than the amplitude of the cyclic variation.
The similarity between  this variability timescale and P$_{flare}$ is interesting,
but we unfortunately do not have sufficient orbital phase coverage to establish
when these cyclical variations are present and when they are not.  There are two
times in Figure 7 when they are absent, one of which ($\phi=$0.258),
intriguingly,  coincides with the conjunction corresponding to M$_1$ in back of
its low-mass companion.    If the optical $\sim$2-hour oscillations were intrinsic to
the neutron star region, this would be the most favorable phase to observe them.  The absence
of  oscillations at this phase is thus perplexing.  

\begin{figure}
\centering
\includegraphics[width=9cm]{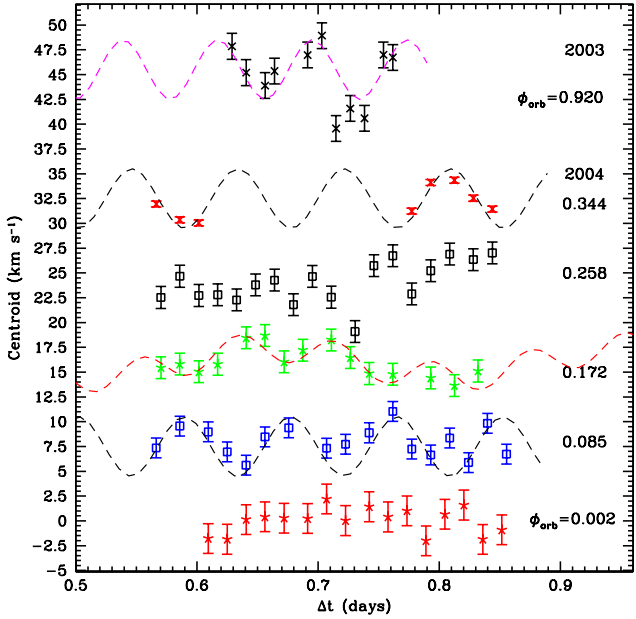}
\caption{Centroids of individual spectra obtained in 2003 and 2004  with the MES-SPM
spectrograph plotted as a function of time, with arbitrary initial times.
Different nights are given a vertical shift for clarity.  Error bars correspond to the
uncertainties determined from the ISM-line measurements.  The sine curves
have frequency of 11.4 cycles day$^{-1}$.
}
\end{figure}

\subsection{Line profile variability}

The large scatter in the RV curve of LSI+65 010 can be traced to line profile
variability, as illustrated in  Figure 8 where the He I 5875 \AA\ line profiles
are stacked in order of increasing phase.  The shape of the observed profiles does not
repeat with orbital phase, nor is there a clear pattern of variability over the
orbital cycle. Some spectra display extraordinarily extended blue wings (e.g., the
1995 data) and some spectra have an extended ``red" emission wing (e.g., orbital
phases 0.002, 0.344, among others).   In contrast, the  line-profile variability 
that is predicted from the tidal oscillations is periodic with orbital phase,
the strongest variations occuring in the phase interval 0.0$-$0.3. This is illustrated
in Figure 9, where we plot the line profiles computed in the  e$=$0.16 model\footnote{The
qualitative nature of the variations for a circular orbit is very similar.}.  The perturbation 
on the line profile first appears at periastron passage on the blue wing and moves towards the red
wing, becoming imperceptible by orbital phase $\sim$0.3.  This pattern repeats  over each
orbital cycle. Hence, the observed behavior of He I 5875 \AA\ indicates that  its
line profile is affected by processes in addition to the surface motions produced by  
the tidal forces. 
     
A clue to the source of additional variability can be found by comparing the ratio
of the perturbed and unperturbed theoretical profiles with the observational analogue.
The left-hand panel of Figure 10 presents the theoretical residuals in the phase interval
0.96$-$1.43, showing the emission ``bumps" and excess absorption features as they move
accross the profile.  These features are present only within the velocity range 
$\pm$100 km s$^{-1}$.  The ratios of observed spectra were constructed using  the 2001 
Jan REOSC ($\phi=$0.304) spectrum in the denominator as a template, due to its apparently 
normal shape (see Figure 1).  Before constructing the ratio, the template was shifted in  
velocity  by amounts corresponding to the RV of the spectrum to be used in the numerator.  
Examples are shown in Figure 10  (right) and compared with theoretical residual spectra
at the corresponding orbital phase.   The observational residuals clearly extend over a
broader velocity range and indicate the presence of actual superposed emission, not just 
the effects caused by ratio-ing line profiles affected by tidal oscillations.  This is not
surprising since we know that H$\alpha$ is in emission and that its line profile is
variable (Margon \& Bradt 1977; Koenigsberger et al. 2003).   It is unfortunate, however, 
that  our high resolution data  do not include H$\alpha$  
preventing us from  searching for a connection between the H$\alpha$ emission and 
this deduced superposed emission. But a common source for both  emissions is likely since the
Gaussian full-width at half maximum intensity of the H$\alpha$ emission in the 2001 Oct spectrum is
280 km s$^{-1}$, which is similar to the width of some of the residual He I 5857 \AA\  profiles.

Whether the emission is produced in the vicinity of the collapsed companion or is associated 
with the B-supergiant can be answered with the radial velocity variations, as displayed in Figure 11.
Although there is a trend for more negative RVs  around $\phi=$0, just 
a few of the observations follow this trend, and the amplitude is only $\sim$70 km s$^{-1}$. 
This argues against the residual emission arising primarily in the vicinity of the
collapsed object where large-amplitude RV variations associated with its orbital motion would be
expected.  Hence, we conclude that the wind close to the B-supergiant is the most likely source  
for the bulk of the superposed emission. The atmospheres of B-supergiants are, in general,
very active and the He I 5875 \AA\ line is reported to undergo large RV changes even
in single stars (Kaufer et al. 2006; Morel et al. 2004), so the superposed emission in
LSI+65 010 is just one more manifestation of the activity.  

In summary, the observed line-profile variability in the He I 5875 \AA\ line exceeds that which 
predicted from the tidal interaction model with input parameters as we have adopted for LSI$+$65~010.
This is caused by  superposed emission which we find to be associated primarily with
the B-supergiant.  The presence and variability of this emission  indicates significant surface 
activity which, however, does not appear to correlate with orbital phase and may thus be 
similar to intrinsic variability that is observed for other B-type supergiants (Morel et al. 2004).

\begin{figure}
\centering
\includegraphics[width=9cm]{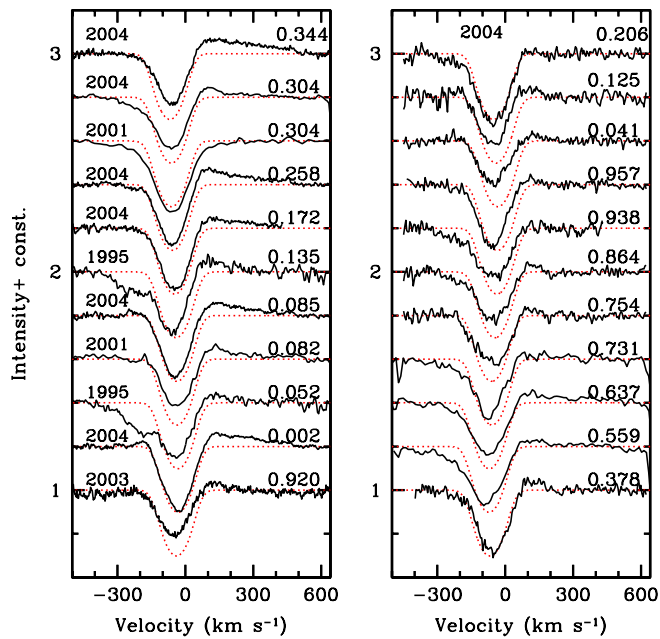}
\caption{Montage of the  He I 5875 \AA\  average (left) line profiles stacked in order of
increasing orbital phase from bottom to top.  The right-hand panel contains line profiles of
nights for which only one or two spectra were acquired. Dotted profiles are the theoretical
line profiles, rotationally broadened to v sin({\it i})=96 km s$^{-1}$, whose velocity shifts
correspond to a Keplerian orbit of a 16 M$_\odot$ star with a 1.7 M$_\odot$ companion.
}
\end{figure}


\begin{figure}
\centering
\includegraphics[width=9cm]{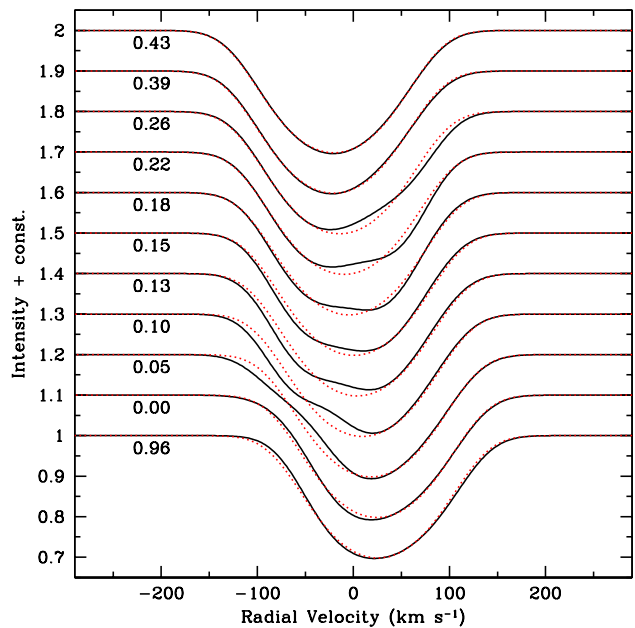}
\caption{Theoretical model line-profiles stacked as a function of orbital
phase for the tidally-perturbed 16 M$_\odot$, R$_1=$25 R$_\odot$ star in the
eccentric orbit.  The dotted profiles correspond to the unperturbed line
profiles.  The strongest variability is observed between periastron
passage ($\phi=$0.00) and $\phi\sim$0.30 and may be described in terms of
a wave pattern that moves from the blue to the red wing.
}
\end{figure}


\begin{figure}
\centering
\includegraphics[width=9cm]{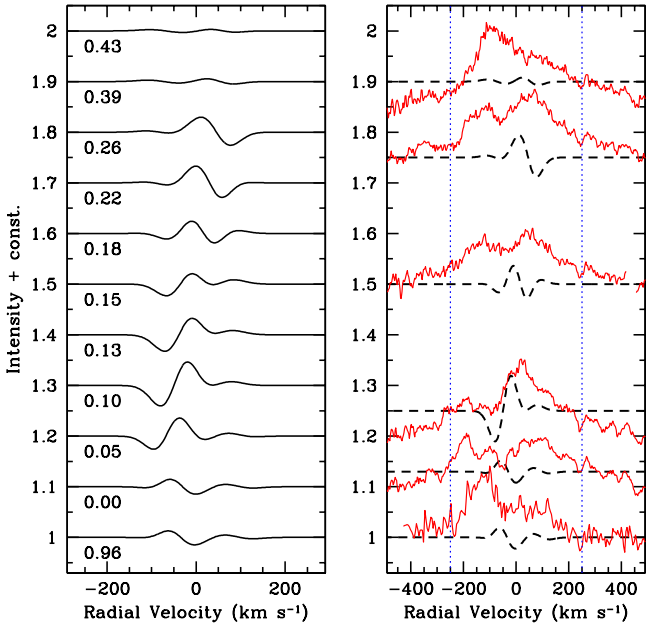}
\caption{Left panel: Ratios of theoretical line profiles of the previous figure,
illustrating the predicted differences between the perturbed  and 
corresponding unperturbed photospheric absorption.  Orbital phases are indicated.  
Right panel: Ratios of the observed profiles  compared to the same theoretical 
variations (dashes) that are illustrated in the right-hand panel. The much
broader velocity range in the observations (indicated by the vertical lines) 
implies the presence of residual  emission, in addition to variability of the
underlying photospheric absorption. 
}
\end{figure}

\begin{figure}
\centering
\includegraphics[width=9cm]{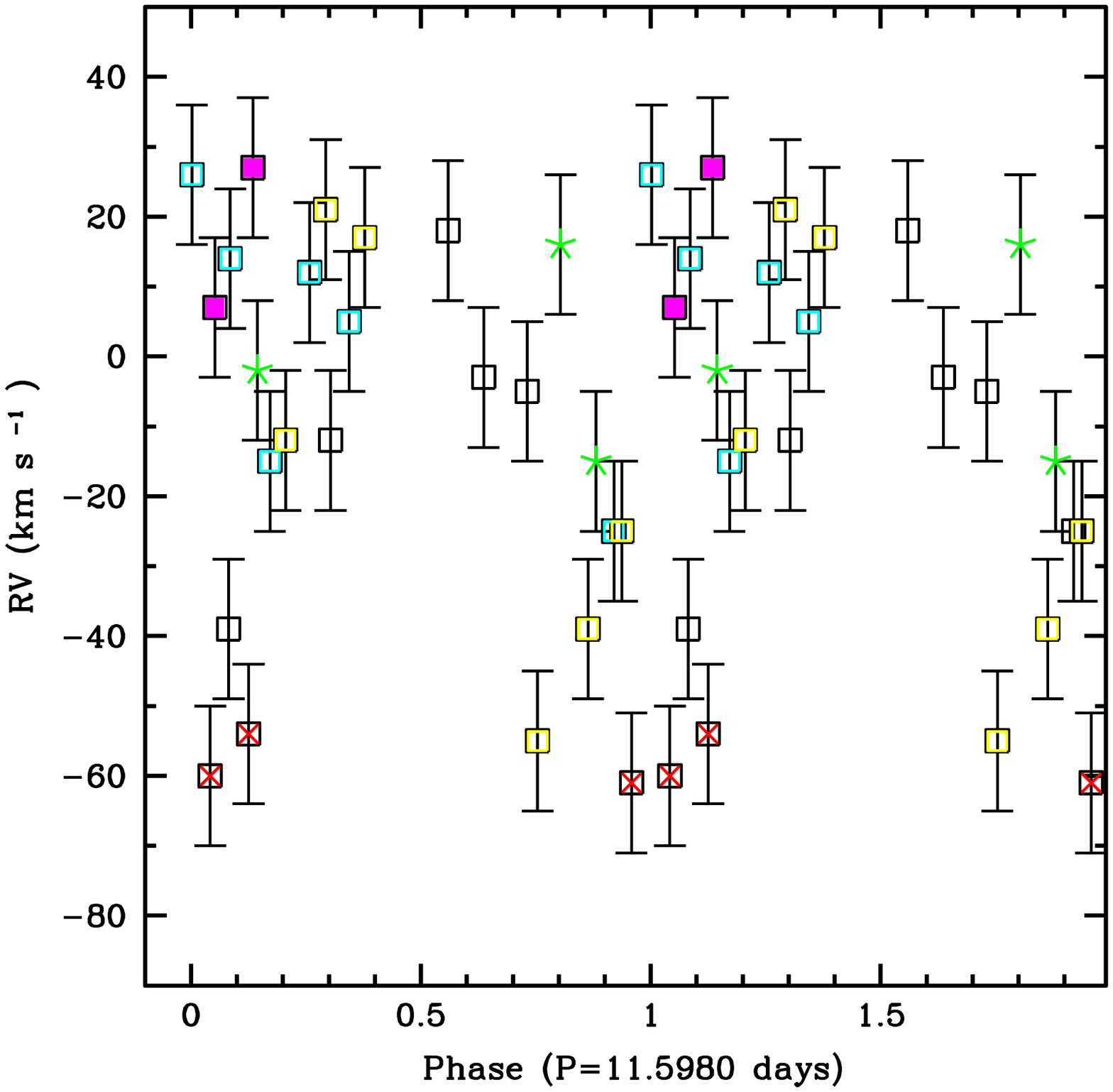}
\caption{Centroid of the residual emission as a function of orbital phase.  Stars represent
the 1993 data; filled-in squares, the 1995 data;  and open squares all other data except
the 2004 Oct values which are indicated with crosses within open squares.
A trend for more negative velocities around $\phi=$0  is present, although the small amplitude
suggests an origin near the B-star rather than near the collapsed companion.
}
\end{figure}

\subsection{The 30.7 day ``superorbital" period}

If the ``superorbital" period discovered by Farrell et al. is related to oscillations of the
primary star, this period should appear in the optical data.  Beskrovnaya (1988)  reports 
polarization variations with a characteristic timescale of $\sim$month, supporting this
possibility.  Although this variability timescale is not apparent in the RV data, we do
find indications of its presence in the strength of the residual emission.
Figure 12 is a plot of the residual emission equivalent width measured by integrating over
the entire emission feature,  showing a possible modulation with P=30.79 days, 
which is within the $\pm$0.1 day uncertainty given by Farrell et al.    
The Farrell et al.  T$_0$ corresponds to ``minimum modulation of the light curve" which,
according to their Fig. 3, occurs during the transition from minimum to maximum X-ray flux.  Hence,
the X-ray P$_{super}$  ``high" state is at phase $\phi_{super} \sim$0.25. This coincides with the
general trend of the data in Figure 12.

The residual emission line profiles are illustrated in Figure 13 where they are  stacked in 
order of increasing 30.79-day phase.  There is a trend  for broader and more symmetrical emission 
near phase 0 (bottom of the plot), with a narrower central peak developing 
thereafter.  At other phases, two peaks are often observed, but the relative intensity 
of the two peaks can change from one night to the next. 

\begin{figure} 
\centering
\includegraphics[width=9cm]{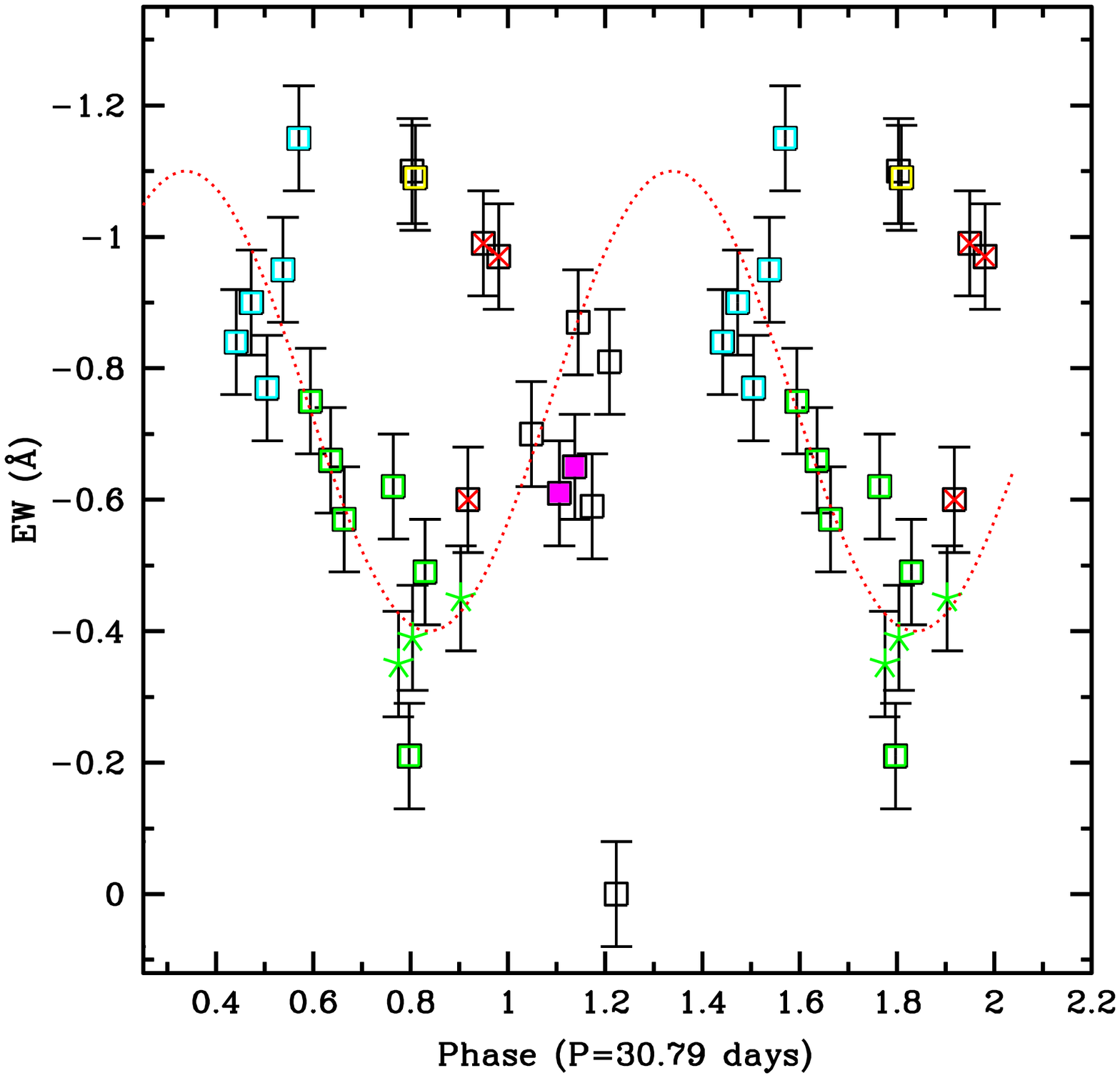}
\caption{Equivalent width of the residual emission  plotted as a function of the
P$_{super}=$30.79 day phase.  A possible modulation is present, illustrated with
a sine curve.  Symbols are the same as in Figure 11.  Here phase 0.0 corresponds to 
the time of minimum modulation in the X-ray light curve of Farrell et al., with X-ray 
``high" occuring near $\phi_{super}=$0.25. 
}
\end{figure}

\begin{figure}
\centering
\includegraphics[width=9cm]{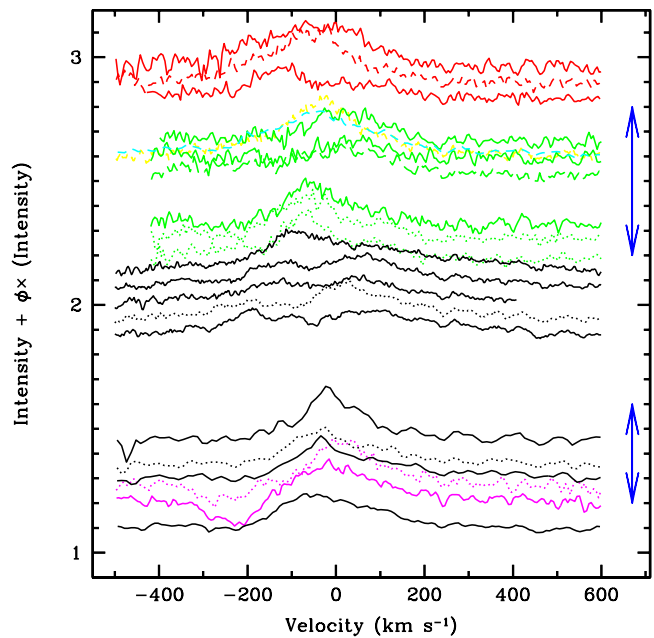}
\caption{He I 5875 \AA\ residual emission stacked in order of increasing P$_{super}=$30.79-day 
phase (from bottom to top), and with separations proportional to the $\phi_{super}$. 
As in Figure 12, zero phase corresponds to ``minimum modulation" or the X-ray light curve 
(Farrell et al.). The arrows  indicate our deduced approximate times of X-ray minimum 
($\phi_{super}\sim$0.6-0.8) and X-ray maximum ($\phi_{super}\sim$0.1-0.3). Dots are used to 
differentiate between overlapping spectra on the plot.
}
\end{figure}

\begin{table*}
\caption{System and  parameters for the eccentric (E) and two circular orbit (C1,C2) models.}
\label{table:2}
\centering
\begin{tabular}{llllll}
\hline\hline
Parameter & Empirical values &Model-E & Model-C1&Model-C2& Comments\\
\hline
P$_{orb}$ (days)    & 11.5995$\pm$0.0053&  11.5980 & 11.5980  & 11.5980   & den Hartog et al. \\
P$_{super}$ (days)    & 30.7        & none     & 30.4 &  30.7& Farrell et al.  \\
P$_{flare}$ (hours)  & $\sim$2.7    &1.6$-$3.4 & $\sim$2.3 &$\sim$3.2 & Hall et al.   \\
$e$                 & 0.0 or 0.16   &  0.16    & 0.00& 0.00 &  Crampton et al.  \\
T$_{ecc}$ (MJD)     & 44134.4       & ---  & --- & --- & Crampton et al. eccentric orbit \\
T$_{circ}$ (MJD)    & 44133.8       & ---  & --- & --- &  Crampton et al. circular orbit \\
T$_{X-orb}$ (MJD)   & 51825.2       & ---  & --- & --- & den Hartog et al. ``X-ray maximum" \\
T$_0$   (MJD)       & 43669.9       &---  & ---&  --- &  T$_{circ}$-40 cycles; elongation \\
T$_{X-super}$ (MJD) & 50108.2       & ---  & ---&---  & Farrell et al. ``minimum F$_X$ modulation"\\
M$_1$ (M$_\odot$)   &  16.0$\pm$5   &   16  &  16  & 16 &  Reig et al.  \\
M$_2$ (M$_\odot$)   &    1.7        &   1.7 & 1.7  & 1.7&   Reig et al. \\
R$_1$  (R$_\odot$)  &    37$\pm$15  &   25  & 40   &36.2& Reig et al.  \\
v sin i (km s$^{-1}$)& 96$\pm$20    &  100  & 96   & 96 & Reig et al.         \\
$i$ (deg)           &  61$-$90      &   90  & 65   & 90 & Reig et al.    \\
v$_{rot}$ (km s$^{-1}$)& 110        &  100  & 106  & 96 &         \\
$\Omega_{per}$ (deg) & ---          &   27  &---   & ---& This paper  \\
$\beta_0$          & 0.4-0.9       &  0.66 & 0.615&0.615& $\omega_0/\Omega_{per}$  \\
$\nu$ (R$_\odot^2$ day$^{-1}$)& --- &  0.087& 0.350&0.120   & viscosity  \\
$\delta$R/R$_1$     & ---          &  0.0067& 0.0050&0.0050 & thickness of external layer  \\
\hline
\end{tabular}
\end{table*}


\begin{figure}
\centering
\includegraphics[width=9cm]{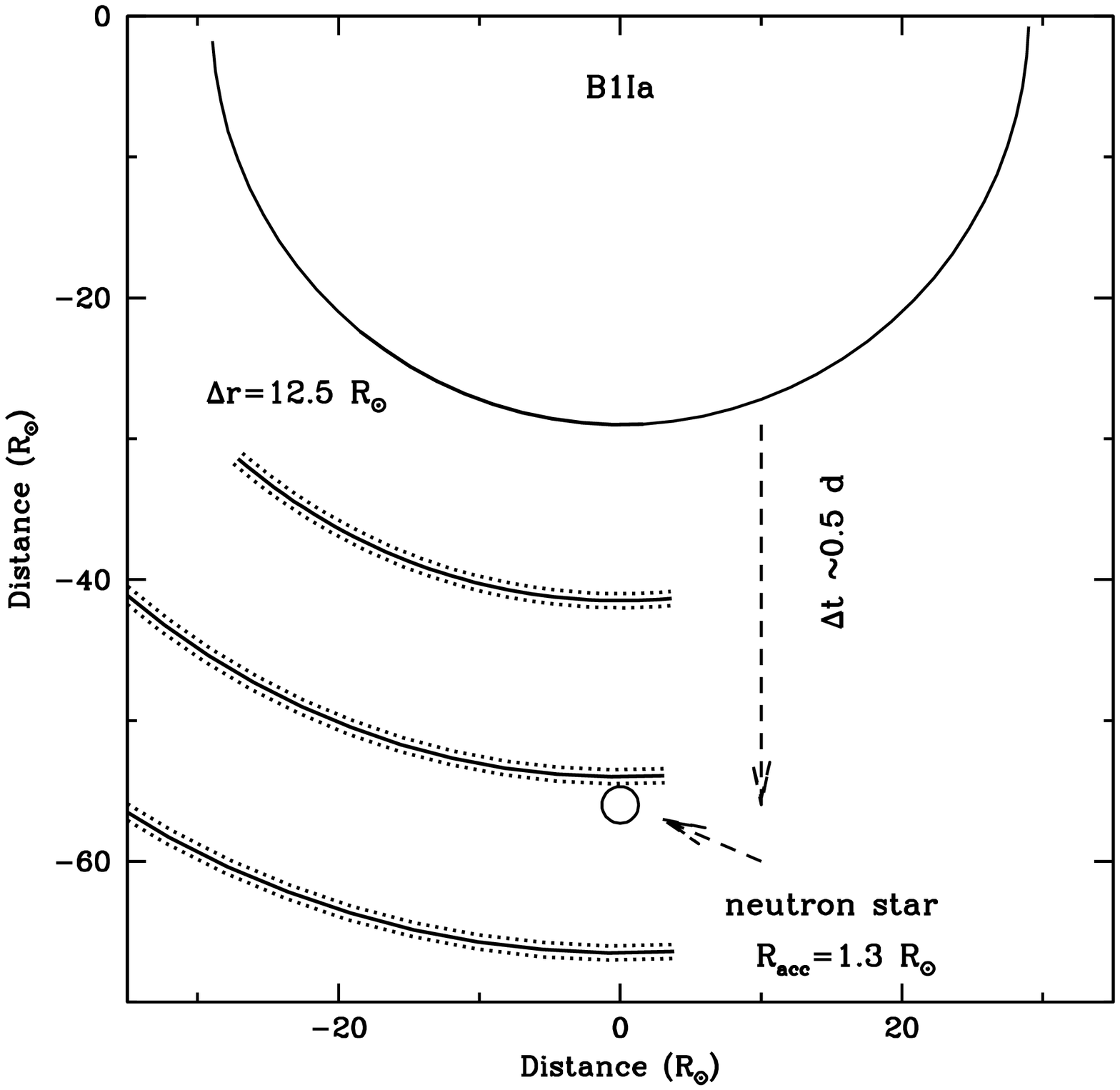}
\caption{Schematic illustration of the scenario for the $\sim$2.7 hour X-ray
periodicity, under the assumption that the oscillations near the tidal bulge
are responsible for modulating the wind of the B-supergiant.}
\end{figure}

\section{Conclusions}

Tidal interactions in non-synchronously rotating binary systems  produce strong perturbations 
on the external layers of the stars, particularly when the orbital separations are 
relatively small and the stellar radii are large. The binary system LSI$+$65 010=2S0114$+$650
is precisely such a system.   The exploratory calculations performed for this paper are aimed 
at establishing the plausibility that the X-ray variability patterns of LSI+65 010$=$2S0114+650 may be 
attributable to these tidal interactions.  We show that  the  oscillations on the surface of
the B-supergiant that are predicted in our calculations have timescales that are consistent 
with the periodic timescales found in X-ray data.  We propose that these oscillations may 
modulate the B-star's wind, leading to a fluctuating accretion rate onto the 
neutron star, and hence, X-ray variability.  
Figure 14 is a cartoon depicting the  scenario for oscillations with
P$\sim$2-3 hours which are assumed to produce enhanced-density shells.  This idea, originally
proposed by Finley et al. (1992) was discarded when the optical component of 2S0114$+$650 was
found to be a B-supergiant instead of a $\beta$ Cep-type star, since such short timescale
pulsations are not predicted for single supergiant stars.   We rescue Finley et al.'s hypothesis by
proposing that the presence of a companion can lead to the appearance of such short periods. 

The precise mechanisms by which the  oscillations may 
translate into a structured stellar wind are not clear, but we speculate that an
answer may lie in the combination of the radial pulsation components and the shear energy 
dissipation due to the tangential motions of the outer stellar layers sliding over the  more 
slowly rotating inner layers.  These two combined sources (the radial pulsation component and the 
shear energy dissipation), when added to the radiation field that normally drives the stellar wind, 
might cause the inhomogeneities to appear on the same timescale as that of the largest amplitude oscillations.  
In order to establish a connection between the tidal oscillations and the modulation of  mass-ejection
in LSI$+$65 010, it is necessary to analyze spectral lines that are less susceptible to contamination
from emission, together with observations of H$\alpha$.  It is important to note that such observations
have been successfully performed in other stars (Fullerton, Gies \& Bolton 1996; Prinja et al. 2004;
Kaufer et al. 2006) showing a link between photospheric activity  and perturbations in the dense inner
regions of stellar winds.

It is interesting to note that the largest oscillation  amplitudes  induced by the tidal 
forces occur near the primary tidal bulge.  Hence, the periodic modulation of the stellar wind may 
only be present in a small solid angle oriented in the general direction of the neutron star.  
The localized nature of the   oscillations would explain the lack of a clear $\sim$2.7 hr periodicity in  
the  He I 5875 \AA\  data, since the irregular variations that are present over the whole B-star 
surface most likely drown  out the small-scale periodic effects.  Thus, an interesting conclusion
is that the X-ray emission is a much finer diagnostic of the local wind structure than the optical 
line profiles, the latter reflecting the integral of the absorption/emission over the stellar disk. 

The tidal interaction model predicts a large difference between the pulsation energy dissipation rates, 
$\dot{E}$, near  periastron and apastron for an eccentric orbit.  This implies a significant variation in 
surface activity as a function of orbital phase, assuming that the surface activity is associated with  
$\dot{E}$.  X-ray  variability on the orbital timescale is indeed present, but the line profiles in our 
optical data do not display strong changes between periastron and apastron phases, weakening the
case for an eccentric orbit.
In a circular orbit, $\dot{E}$ remains constant over orbital phase and thus no modulation
of the stellar wind with P$_{orb}$  is expected.  If the orbit of LSI$+$65 010 is circular, the 
observed X-ray variability is likely due to eclipse effects, rather than to differences in the 
accretion rates onto the companion.  A circular orbit would also be consistent with the existence 
of a ``superorbital" period.

Our final consideration concerns the possibility that part of the X-ray emission may actually
originate in the B-supergiant.   In particular, the tangential component 
of the perturbation produces a differentially rotating  structure in the surface layers that one 
might speculate could lead to magnetic field generation.  Admittedly, the timescales over which 
a particular region of the star develops this structure  may be too short for  
magnetic field amplification mechanisms for massive stars, such as proposed by Spruit (2002),  
to operate. However, it would be interesting to  explore the  mechanisms that could convert tidal 
energy into magnetic field energy, as has been done  for low-mass stars (for example, Zaqarashvili 
et al. 2002).  Furthermore, observational evidence for  surface magnetic field structures 
in OB stars now exists (Henrichs, Neiner, Geers 2003), so it would be interesting also to explore the
effects of the tidal oscillations on assumed pre-existing  magnetic field structures 
and determine the X-ray emitting characteristics due to these perturbations.


\begin{acknowledgements}
We thank Jos\'e Alberto L\'opez and Hortensia Riesgo for their participation in the
2004 Nov. and Dec. observations, and an anonymous referee for suggestions and comments
that helped to improve the manuscript.
This work was supported by CONACYT grants 42809-E, 43121-E  and UNAM/PAPIIT grants IN112103,
IN108406-2, IN108506-2, and IN119218.
\end{acknowledgements}

\centering
%

\centering
%

\end{document}